\documentclass[12]{iopart}

\usepackage{iopams}
\usepackage{graphicx}
\usepackage{subfigure}
\usepackage{iopams}
\usepackage{bbold}

\begin{document}

\title{Proof-of-Concept of Real-World Quantum Key Distribution with Quantum Frames}

\author{I. Lucio-Martinez$^1$, P. Chan$^2$, X. Mo$^1$, S. Hosier$^3$, W. Tittel$^1$}

\address{$^1$University of Calgary, Institute for Quantum Information Science and Department of Physics and Astronomy\\
$^2$University of Calgary, Advanced Technology Information Processing Systems Laboratory and Department of Electrical and Computer Engineering\\
$^3$Southern Alberta Institute of Technology}

\begin{abstract} We propose a fibre-based quantum key distribution system, which employs polarization qubits encoded into faint laser pulses. As a novel feature, it allows sending of classical framing information via sequences of strong laser pulses that precede the quantum data.
This allows synchronization, sender and receiver identification, and compensation of time-varying birefringence in the communication channel. In addition, this method also provides a platform to communicate implementation specific information such as encoding and protocol in view of future optical quantum networks. We demonstrate in a long-term (37 hour) proof-of-principle study that polarization information encoded in the classical control frames can indeed be used to stabilize unwanted qubit transformation in the quantum channel. All optical elements in our setup can be operated at Gbps rates, which is a first requirement for a future system delivering secret keys at Mbps. In order to remove another bottleneck towards a high rate system, we investigate forward error correction based on Low-Density Parity-Check Codes.
\end{abstract}

\maketitle

\section{Introduction\label{sec1}}

Based on the particular properties of single quantum systems, quantum key distribution (QKD) promises cryptographic key exchange over an untrusted, authenticated public communication channel with information theoretic security \cite{Bennett1984,Shor2000}. Significant academic \cite{Gisin2002,Scarani2008a}, and industrial effort \cite{Industrial_QKD} has been devoted to the development of point-to-point (P2P) QKD systems based on attenuated laser pulses or entangled photons, and the first fully functional prototype of a quantum cryptographic network consisting of pre-established P2P links in a trusted node scenario has recently been demonstrated \cite{SECOQC} (see also \cite{DARPA2005}). Furthermore, various proof-of-principle demonstrations of quantum teleportation and quantum memory (see \cite{Tittel2008, Hammerer2008} and references therein) have been reported, which will eventually allow building of fully quantum enabled networks \cite{Gisin2007,Kimble2008}, e.g. for perfectly secure communication in settings with un-trusted nodes and over large distances \cite{Briegel1998,Duan2001}.

Despite these remarkable achievements, the building of a reconfigurable real-world QKD network still requires significant progress, even when limiting quantum communication to qubits encoded into faint laser pulses and to entangled qubits. Among the issues to be solved is the necessity to route  quantum data from any sender to any receiver. The possibility to use active optical switches to send quantum information to different users has first been demonstrated in 2003 \cite{Tolliver2003}. However, the question regarding the addition of sender and receiver addresses to the quantum data (which is not required in pre-established P2P links) has, to the best of our knowledge, never been addressed. Beyond routing, another requirement for quantum networks is path stabilization between sender and receiver, i.e. to ensure that carriers of qubits prepared at Alice's arrive unperturbed at Bob's. This includes control of the properties of the quantum channel, e.g. birefringence in an optical fibre, and the establishment of a common reference frame at Alice's and Bob's, e.g. a direction or a precise time-difference, depending on the property chosen to encode the qubit \cite{Tittel2001}. Current P2P QKD systems are either of the 'plug \& play' type and automatically stabilize the quantum channel \cite{Muller1997,Zbinden1997}, or achieve unperturbed quantum communication by adding from time to time short sequences of classical control information \cite{Yuan2005}. However, neither method allows communication of the properties that are important in reconfigurable networks, including  sender and receiver address, or the specific QKD protocol or type of qubit encoding chosen\footnote{Note that this information can also be sent through another (classical) channel. However, given that control information for channel stabilization has to be sent in any case (except for auto-compensating systems such as the 'plug \& play' system), it is natural to consider sending the network relevant control information through the quantum channel as well.}.

In this article we propose the use of quantum frames as a flexible framework for sensing, communicating and controlling the parameters relevant in a QKD network setting. Our approach is sufficiently flexible to accommodate for current and future quantum technology or applications, including technology from different vendors, which is important in view of open quantum networks. We demonstrate the suitability of our solution for quantum key distribution with polarization qubits over a 12 km real-world fibre optic link.

This article is organized as follows: In section \ref{sec2} we present the general idea of quantum frames. We then discuss the principle QKD setup (section \ref{sec3}), and give further details of key components (section \ref{sec4}). After presenting the properties of our fibre optics link (section \ref{sec5}), we describe the QKD field tests and discuss the results (section \ref{sec6}), and then elaborate briefly on some issues related to the security of the key establishment (section \ref{sec7}). In section \ref{sec8} we present the status of our classical post processing, required to distill a secret key, specifically the possibility of hardware implementation of one-way error correction. We present our conclusions in section \ref{sec9}.

\section{Quantum Frames\label{sec2}}

\begin{figure}
\begin{center}
  \includegraphics[width=11cm]{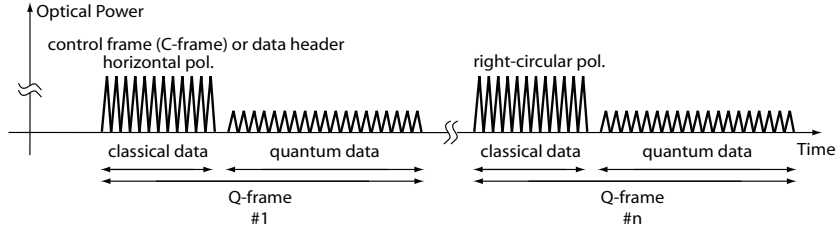}
  \caption{Quantum framing with alternating classical control frames (C-frames, inspired by the Ethernet protocol) and quantum data. In the here reported implementation, subsequent C-frames encode different polarization states (horizontal, vertical and circular), each one used to independently stabilize one particular set of polarization qubit basis states. \label{fig2_1}}
\end{center}
\end{figure}

To add control functionalities to the communication between Alice and Bob, we propose supplementing the quantum data (e.g. qubits) with classical control frames. The control frames (C-frame), encoded into strong laser pulses, alternate with the quantum data, and a pair of classical/quantum data forms a quantum frame (Q-frame), see \fref{fig2_1}. The C-frame allows synchronizing sender Alice and receiver Bob, facilitates time-tagging, and provides a platform to communicate sender and receiver address (for routing or packet switching) plus implementation specific information such as encoding (e.g. polarization or time-bin qubit \cite{Tittel2001}) and protocol (e.g. BB84 \cite{Bennett1984}, decoy state \cite{Hwang2003,Ma2005,Wang2005}, or B92 \cite{Bennett1992}). This is interesting in view of open, reconfigurable networks comprising different QKD technologies.

The classical information in our implementation is encoded into specific polarization states, allowing assessment and compensation of time-varying birefringence in the quantum channel. Note that the compensation scheme can easily be adapted to other QKD setups employing e.g. time-bin qubits, entanglement, or quantum repeaters. Furthermore, the C-frames can be used to asses channel loss, which may be important for routing.

\section{Our QKD System \label{sec3}}

\begin{figure}
  \begin{center}
    \includegraphics[width=12cm]{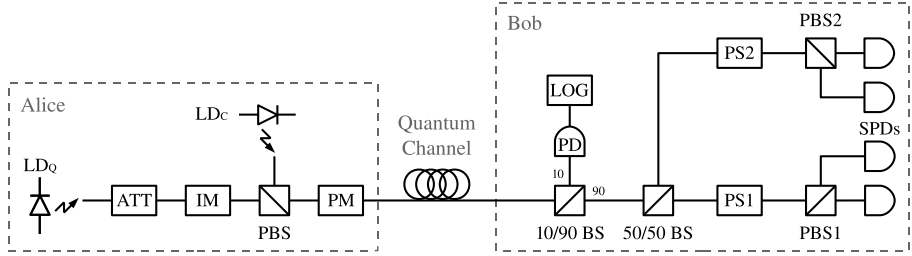}
    \caption{Schematic of our QKD system. \label{fig3_1}}
  \end{center}
\end{figure}

Our QKD system is based on polarization qubits, and employs the BB84 protocol~\cite{Bennett1984}, supplemented with two decoy states~\cite{Hwang2003,Ma2005,Wang2005}. It allows alternating sequences of strong and faint laser pulses, encoding classical data and quantum data, respectively. A simplified schematic of the QKD system is depicted in \fref{fig3_1}. Alice uses two laser diodes to generate the classical data (LD$_\mathrm{C}$) and the quantum data (LD$_\mathrm{Q}$). The pulses emitted from LD$_\mathrm{Q}$ are first attenuated by an optical attenuator (ATT), and then sent through an intensity modulator (IM) to create signal and decoy states with different mean photon numbers. To create vacuum decoy states, no electrical pulses are sent to LD$_\mathrm{Q}$. The horizontally polarized faint pulses are then transmitted through a polarization beam splitter (PBS), and combined with the strong, vertically polarized pulses from LD$_\mathrm{C}$. All pulses are then sent to a polarization modulator (PM), where horizontal (H), vertical (V), right (R), or left (L) circular polarization states can be created.

Quantum and classical data is transmitted to Bob through a quantum channel. At Bob's end, 10\% of the light is directed towards a fast photo detector (PD) followed by a logic device (LOG). The detector and the logic device, which were not implemented in our investigation, will read the information encoded in the classical data and take appropriate action, e.g. for clock synchronization, optical routing, or communication of protocol specific information used by Bob for the measurement and subsequent processing of the quantum data.

The remaining light is split at a 50/50 beam splitter (BS), and directed to two polarization stabilizers (PS1, PS2) followed by polarization beam splitters (PBS1, PBS2) and single photon detectors (SPDs). PS1 ensures that horizontally polarized classical data, and hence qubits, emitted at Alice's arrive unchanged at PBS1. Similarly, PS2 is set up such that right circular polarized classical data and qubits emitted at Alice's always impinge horizontally polarized on PBS2. Since the transformation in the quantum channel is described by a unitary matrix (i.e. orthogonal states remain orthogonal), our stabilization scheme ensures that qubits prepared in H and V, or R and L states arrive horizontally and vertically polarized on PBS1 or PBS2, respectively. Hence, the two sets of PS, PBS and two SPDs both allow compensation of unwanted polarization transformations in the quantum channel, and projection measurements onto H, V, R and L, as required in the BB84 protocol.  Note that our scheme does not prevent H and V created at Alice's from arriving in an arbitrary superposition of H and V at PBS2 (similar for R and L at PBS1). However, these cases do not cause errors as they are eliminated during key sifting.

\section{Polarization and Intensity Modulators \label{sec4}}

\begin{figure}
  \begin{center}
    \subfigure[]
      {
        \includegraphics[height=1.8cm]{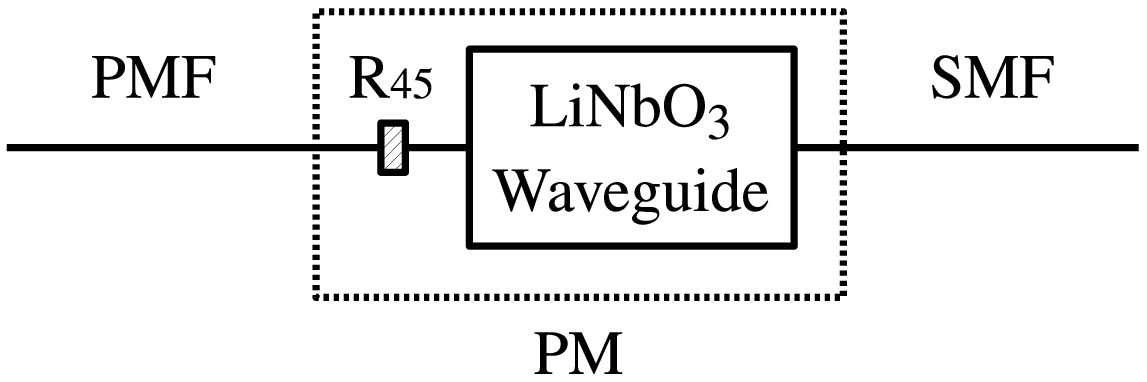}
        \label{fig4_1a}
      }
    \subfigure[]
      {
        \includegraphics[height=1.8cm]{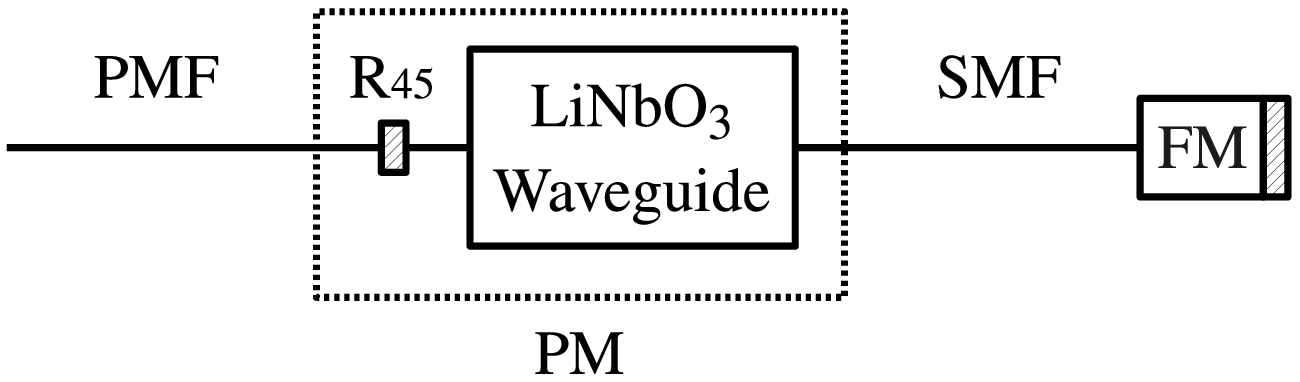}
        \label{fig4_1b}
      }
    \subfigure[]
      {
        \includegraphics[height=1.8cm]{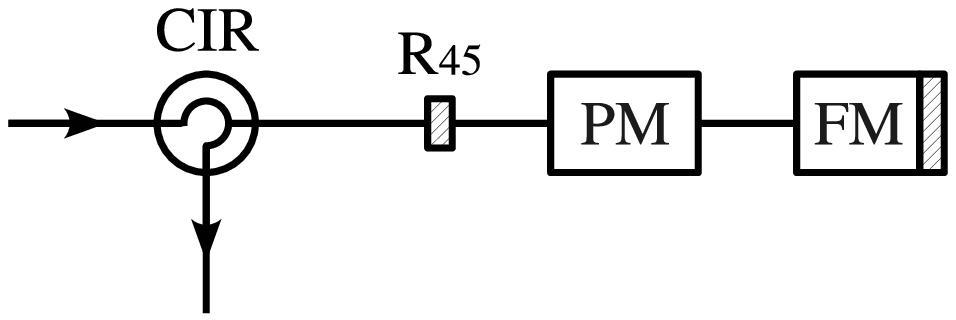}
        \label{fig4_1c}
      }
    \subfigure[]
      {
        \includegraphics[height=1.8cm]{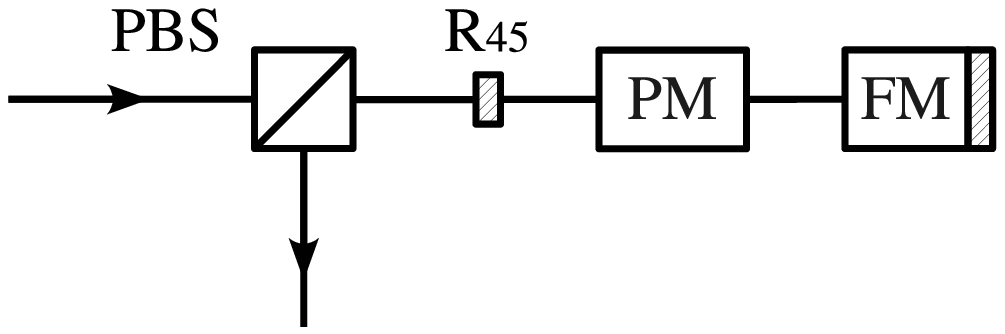}
        \label{fig4_1d}
      }
    \caption{Schematics of (a) one-way polarization modulator, (b) basic unit, (c) two-way polarization modulator based on basic unit, (d) two-way intensity modulator based on basic unit.}
  \end{center}
\end{figure}

\subsection{One-Way Polarization and Intensity Modulator }

Initially, we used a commercial LiNbO$_\mathrm{3}$ phase modulator (PM) and a Mach-Zehnder intensity modulator in a one-way configuration to achieve fast polarization and intensity modulation. \Fref{fig4_1a} shows the schematics of the polarization modulator, i.e. a phase modulator with polarization maintaining input fibre (PMF) whose slow axis is rotated 45 degrees (R$_\mathrm{45}$) with respect to the optical axis of the modulator waveguide, and standard single mode output fibre (SMF). Hence, horizontally polarized input light, which propagates parallel to the slow axis of the PMF, is split into two components, where each one propagates along one axis of the waveguide. By applying a control voltage to the phase modulator, a phase shift is introduced between the two components, resulting in a polarization modulation.

Unfortunately, the phase modulator features significant polarization mode dispersion (PMD) for 500 ps long optical pulses resulting in a polarization extinction ratio (PER), i.e. the ratio between optical power in two orthogonal polarization states, of only 16 dB. Moreover, we found both the phase and intensity modulator to be temperature sensitive -- a change of environmental temperature or heating caused by passing a current through the impedance matching resistance inside the modulators causes a variation of the polarization state, or the intensity level, of the output light. This would have a direct impact on the quantum bit error rate (QBER) and stability of our QKD system.

\subsection{The ``Basic Unit"}

To overcome these problems, we designed a ``basic unit" (see \fref{fig4_1b}) consisting of a phase modulator (PM) with 45 degree rotated input PMF and a Faraday mirror (FM) \cite{Faradaymirror}. As explained below, this allows building stable polarization and intensity modulators by means of a go-and-return configuration (the light travels twice and in orthogonal polarization states through the phase modulator).

To explain how the basic unit works, we calculate the polarization evolution of light using Jones calculus:

\begin{equation}
  \textbf{J}_\mathrm{out}
  = M_\mathrm{BU} \cdot \textbf{J}_\mathrm{in}.
  \label{eq4_1}
\end{equation}

\noindent  $\textbf{J}_\mathrm{in}$ and $\textbf{J}_\mathrm{out}$ denote the Jones polarization vectors of the input and output light, respectively, and $M_\mathrm{BU}$ is the polarization transformation matrix of the basic unit:

\begin{equation}
  \eqalign{
   M_\mathrm{BU} = \overleftarrow{M}_{\mathrm{PMF}} \cdot {R}^{\dagger}_{\mathrm{45}} \cdot \overleftarrow{M}_{\mathrm{WG}} & \cdot \overleftarrow{M}_{\mathrm{SMF}} \cdot FM \\ & \cdot \overrightarrow{M}_{\mathrm{SMF}} \cdot \overrightarrow{M}_{\mathrm{WG}} \cdot {R}_{\mathrm{45}} \cdot \overrightarrow{M}_{\mathrm{PMF}}.
  }
  \label{eq4_2}
\end{equation}

\noindent $M_\mathrm{SMF}$, $M_\mathrm{PMF}$, and $M_\mathrm{WG}$ denote the polarization transformation matrices of the single mode fibre, the polarization maintaining fibre and the waveguide, respectively, and the arrows on top of the matrices specify the direction of light propagation. $FM$ denotes the effect of the Faraday mirror, and $R_\mathrm{45}$ characterizes the rotation between the polarization maintaining fibre and the waveguide. Assuming that one can neglect all temperature or mechanical stress mediated changes of the properties of the fibres and the waveguide between two subsequent passages of a pulse of light (around ten nanoseconds in our setup), and that these elements do not feature polarization dependent loss, we have

\begin{equation}
  \eqalign{
    \overleftarrow{M}_{\mathrm{PMF}} = {M}^{\dagger}_{\mathrm{PMF}}, ~~  \overrightarrow{M}_{\mathrm{PMF}} = {M}_{\mathrm{PMF}}, \\
    {M}_{\mathrm{PMF}} =
    \left[
      \begin{array}{cc}
        1 &  0 \\
        0 & e^{i\phi_{\mathrm{PMF}}}\\
      \end{array}
      \right], \\
  }
\end{equation}

\noindent where $M^{\dagger}$ stands for the adjoint matrix of $M$, and $\phi_{\mathrm{PMF}}$ is the phase shift caused by the birefringence of the polarization maintaining fibre. Furthermore, we have

\begin{equation}
  \eqalign{
    \overleftarrow{M}_{\mathrm{SMF}} = {M}^{\dagger}_{\mathrm{SMF}}, ~~ \overrightarrow{M}_{\mathrm{SMF}} = {M}_{\mathrm{SMF}}, \\
    M_\mathrm{SMF} = \left[
      \begin{array}{cc}
        \sqrt{a} &  \sqrt{1-a}e^{i\alpha} \\
        \sqrt{1-a}e^{i\beta} & -\sqrt{a}e^{i(\alpha+\beta)}\\
      \end{array}
      \right], \\
  }
\end{equation}

\noindent where $M_{\mathrm{SMF}}$ is the most general unitary matrix describing polarization transformations. The matrices of the waveguide are given by

\begin{equation}
  \overrightarrow{M}_{\mathrm{WG}} =
  \left[
  \begin{array}{cc}
    1 &  0 \\
    0 & e^{i(\phi^{\mathrm{in}}_{\mathrm{m}} + \phi^{~}_{\mathrm{e}})}\\
  \end{array}
  \right], ~~
  \overleftarrow{M}_{\mathrm{WG}} =
  \left[
  \begin{array}{cc}
    1 &  0 \\
    0 & e^{-i(\phi^{\mathrm{out}}_{\mathrm{m}} + \phi^{~}_{\mathrm{e}})}\\
  \end{array}
  \right],
\end{equation}

\noindent where $\phi^{\mathrm{in}}_{\mathrm{m}}$ and $\phi^{\mathrm{out}}_{\mathrm{m}}$ denote the phase shifts during the two subsequent passages of the light through the waveguide, as determined by the modulation voltage applied to the waveguide, and $\phi^{~}_{\mathrm{e}}$ refers to an additional, wavelength and polarization dependent phase shift (leading to PMD).

The effect of the Faraday mirror is to transform the polarization state of an arbitrary input state of light $\textbf{J}_\mathrm{in}$ with components $j_\mathrm{1},j_\mathrm{2}$
into the orthogonal state \cite{Gisin2002}:

\begin{equation}
  FM\cdot \textbf{J}_{in}=FM \cdot \left[
    \begin{array}{c}
      j_1 \\
      j_2 \\
    \end{array}
    \right]
  =
  \left[
  \begin{array}{c}
    j_2^{*} \\
    -j_1^{*} \\
  \end{array}
  \right]\equiv \textbf{J}_{in}^\perp.
  \label {eq4_fm}
\end{equation}

\noindent
Hence, from equation \eref{eq4_fm}, we obtain the identity

\begin{equation}
  \eqalign{
    FM \cdot M \cdot
    \textbf{J}_{in}
    & = FM \cdot \left[
      \begin{array}{cc}
        A & B \\
        C & D \\
      \end{array}
    \right] \cdot
    \textbf{J}_{in} \\
        & =
    \left[
      \begin{array}{cc}
        D^{*} & -C^{*} \\
        -B^{*} & A^{*} \\
      \end{array}
      \right] \cdot FM \cdot
    \textbf{J}_{in}
  }
\end{equation}

\noindent and thus

\begin{equation}
  \eqalign{
    M^{\dagger} \cdot FM \cdot M \cdot
    \textbf{J}_{in} & =
    \left[
      \begin{array}{cc}
        A & B \\
        C & D \\
      \end{array}
      \right]^{\dagger}
    \cdot FM \cdot \left[
      \begin{array}{cc}
        A & B \\
        C & D \\
      \end{array}
      \right] \cdot
    \textbf{J}_{in}\\
    & = \left[
      \begin{array}{cc}
        A^{*} & C^{*} \\
        B^{*} & D^{*} \\
      \end{array}
      \right]
    \cdot \left[
      \begin{array}{cc}
        D^{*} & -C^{*} \\
        -B^{*} & A^{*} \\
      \end{array}
      \right] \cdot FM \cdot
    \textbf{J}_{in}\\
    & = (A^{*}D^{*}-B^{*}C^{*}) \cdot \mathbb{1} \cdot FM \cdot
    \textbf{J}_{in}\\
    & = det(M^{*}) \cdot
    \textbf{J}_{in}^\perp,
  }
  \label {eq4_mfmm}
\end{equation}

\noindent where $M$ is an arbitrary two-by-two matrix, which may describe wavelength dependent polarization rotations or polarization dependent loss, and $\mathbb{1}$ is the two-by-two identity matrix. Equation \eref{eq4_mfmm} shows that any polarization transformation is compensated by the Faraday mirror; the output polarization state $\textbf{J}_{out}$ is always orthogonal to the input state $\textbf{J}_{in}$, regardless of $M$.

Calculating the product of all matrices in equation \eref{eq4_2}, we obtain

\begin{equation}
  \eqalign{
    M_\mathrm{BU} ~ = & ~ e^{-i( \phi_\mathrm{SMF} + \phi_\mathrm{PMF} + \phi^{~}_\mathrm{e} + \phi^{\mathrm{'}}_\mathrm{m} )} \\
     & \cdot \left[
      \begin{array}{cc}
        \cos{\Delta\phi_\mathrm{m}} & -i e^{i\phi_\mathrm{PMF}} \sin{\Delta\phi_\mathrm{m}} \\
        -i e^{-i \phi_\mathrm{PMF}} \sin{\Delta\phi_\mathrm{m}} & \cos{\Delta\phi_\mathrm{m}} \\
      \end{array}
    \right] \cdot FM, \\
  }
\end{equation}

\noindent where $\phi^{\mathrm{'}}_\mathrm{m} = \frac{ \phi^\mathrm{in}_\mathrm{m}+\phi^\mathrm{out}_\mathrm{m} }{2}$, $\Delta\phi_\mathrm{m} = \frac{ \phi^\mathrm{out}_\mathrm{m}-\phi^\mathrm{in}_\mathrm{m} }{2}$, and $\phi_{\mathrm{SMF}} = \pi-\alpha-\beta$. Accordingly, for a horizontal input state, we find

\begin{equation}
\eqalign{
    \textbf{J}_\mathrm{out}  & = M_\mathrm{BU} \cdot \left[
    \begin{array}{c}
      1 \\
      0 \\
    \end{array}
    \right]\\
    & = e^{-i( \phi_\mathrm{SMF} + \phi_\mathrm{PMF} + \phi^{~}_\mathrm{e} + \phi^{\mathrm{'}}_\mathrm{m} )} \cdot \left[
      \begin{array}{c}
        -i e^{i\phi_\mathrm{PMF}} \sin{\Delta\phi_\mathrm{m}} \\
        \cos{\Delta\phi_\mathrm{m}} \\
      \end{array}
      \right] \\
      & = e^{-i( \phi_\mathrm{SMF} + \phi_\mathrm{PMF} + \phi^{~}_\mathrm{e} + \phi^{\mathrm{'}}_\mathrm{m} )} \cdot \left[
      \begin{array}{cc}
        e^{i\phi_\mathrm{PMF}} & 0\\
        0 & 1
      \end{array}
      \right]
    \cdot \left[
      \begin{array}{c}
        -i  \sin{\Delta\phi_\mathrm{m}} \\
        \cos{\Delta\phi_\mathrm{m}} \\
      \end{array}
      \right].
    \label{eq4_eout1}
}
\end{equation}

\noindent Hence, owing to the use of a Faraday mirror, the polarization and wavelength dependent phase shift $\phi_e$ introduced by the waveguide impacts now on the global phase but does not lead to polarization mode dispersion any more. Furthermore, all (slow) modifications of the polarization modulation due to changes in temperature or mechanical stress of the SM and PM fibres are automatically compensated. The output polarization state thus only depends on the modulation of the waveguide ($\Delta\phi_\mathrm{m}$) and the phase shift induced by the polarization maintaining fibre ($\phi_\mathrm{PMF}$).

\subsection{Two-Way Polarization Modulator}

 We complemented the basic unit to a polarization modulator by preceding it by a polarization maintaining circulator (CIR) that allows separating the input and output optical pulses (see \fref{fig4_1c}). By applying appropriate, short voltage pulses, which are synchronized with the propagations of the optical pulse, to the phase modulator, we can generate horizontal ($\Delta\phi_\mathrm{m}=\pi /2$), vertical ($\Delta\phi_\mathrm{m}=0$), right-hand ($\Delta\phi_\mathrm{m}=-\pi /4$), or left-hand circular polarization ($\Delta\phi_\mathrm{m}=\pi /4$) states. We point out that the existence of the phase introduced by the PM fibre, $\phi_\mathrm{PMF}$, makes circular polarization states unstable. However, note that the four generated polarization states always form two mutually unbiased bases, regardless the value of this phase, as required for secure QKD. Furthermore, as the change in the polarization maintaining fibre is slow, it can be compensated by a polarization stabilizer at Bob's, allowing for the establishment of a sifted key with a small quantum bit error rate (QBER).

\begin{figure}[ht]
  \begin{center}
    \includegraphics{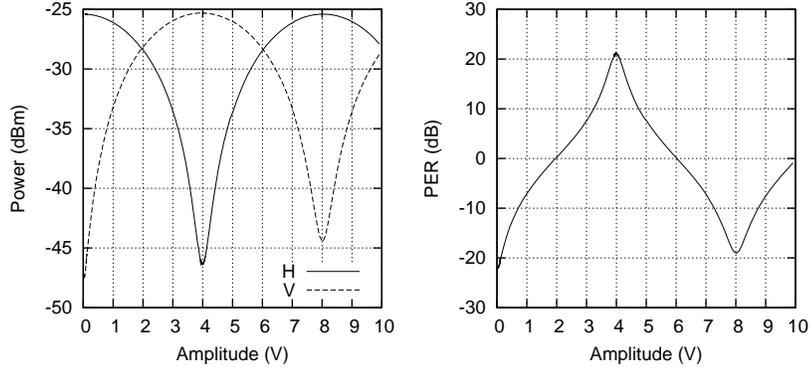}
    \caption{Test of the two-way polarization modulator. In the experiment, the light exiting the modulator was split by a polarization beam splitter (PBS) and the power was measured at the two outputs (H and V) as a function of the modulation voltage. The polarization extinction ratio (PER) is defined as the ratio between the power in the two outputs. \label{fig4_2}}
  \end{center}
\end{figure}

We obtained a polarization extinction ratio of 20 dB for horizontal and vertical polarization states (limited by the light source used to test the polarization modulator), see \fref{fig4_2}, and of 15 dB for left and right circular polarization. We believe the reduced ratio to be caused by state dependent polarization mode dispersion in the circulator, which will be replaced in the near future.


\subsection{Two-Way Intensity Modulator}

Similarly, we built an intensity modulator by preceding the basic unit by a PBS, as shown in \fref{fig4_1d}. The PBS reflects the vertical component of the impinging light. Hence, by varying the polarization state of the light at the output of the basic unit, we can vary the intensity of the vertical
component at the output of the PBS.

\begin{figure}[ht]
  \begin{center}
    \subfigure[]
      {
        \includegraphics{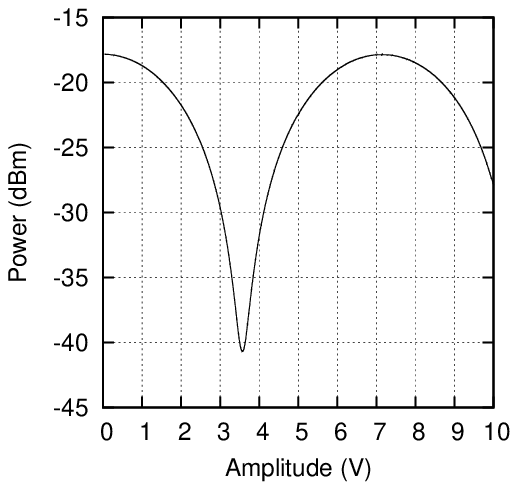}
        \label{fig4_3a}
      }
    \subfigure[]
      {
        \includegraphics{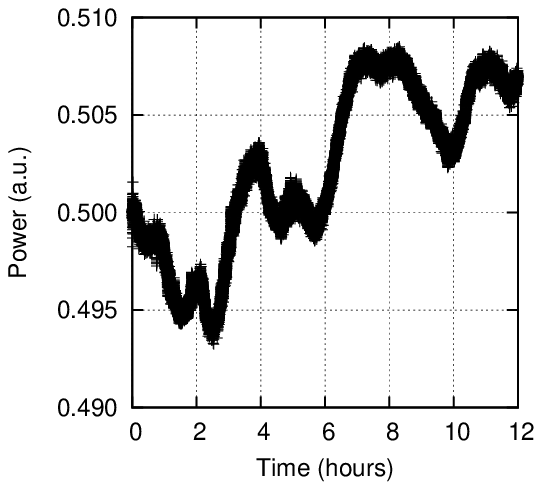}
        \label{fig4_3b}
      }
    \caption{Tests of the two-way intensity modulator. Figure (a) shows the output power as a function of the applied voltage pulse to the phase modulator. The modulator features an extinction ratio of 23 dB. Figure (b) depicts the output power as a function of time. For this measurement, the output power was set to 50\% of its maximum value. The total variation in 12 hours is less than $\pm1.5\%$. This is mostly determined by the power fluctuations of the laser diode, which we found to be $\pm1.15\%$ in 3 hours (note that the latter can be further reduced using external power control).}
  \end{center}
\end{figure}

The intensity extinction ratio, i.e. the ratio between the maximum and minimum intensity at the output of the PBS, exceeds 20 dB (see \fref{fig4_3a}). Moreover, as the phase, $\phi_\mathrm{PMF}$, does not impact on the output intensity, our modulator features an outstanding stability, as depicted in \fref{fig4_3b}. This is important when implementing a decoy state QKD protocol, which relies on accurate preparation of average photon numbers per faint laser pulse.

\section{The Fiber Link\label{sec5}}

\subsection{Loss\label{QChannel_loss}}

\begin{figure}
  \begin{center}
      \includegraphics[width=10cm]{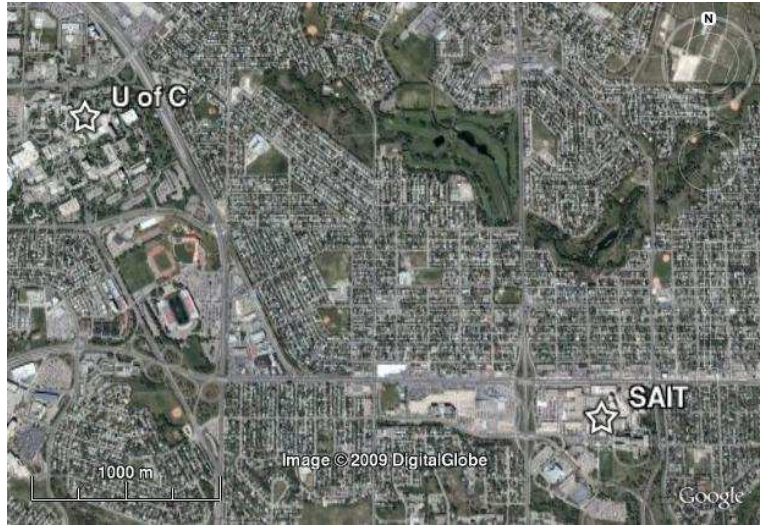}
        \caption[Caption short.]{Satellite view of Calgary, showing the University of Calgary (U of C) and the Southern Alberta Institute of Technology (SAIT).}
        \label{fig5_1}
    \end{center}
\end{figure}

\begin{figure}
    \begin{center}
            \includegraphics{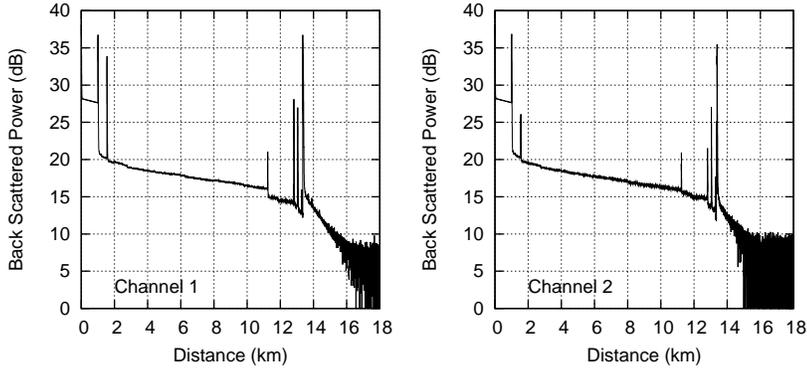}
        \caption[Caption short.]{OTDR traces of the installed fibres. The horizontal axis denotes the distance measured from the laboratory at SAIT. The vertical axis denotes the logarithm of the ratio between the back scattered power detected by the OTDR and a reference power set by the instrument, where a higher value corresponds to more reflected power.}
        \label{fig5_2}
    \end{center}
\end{figure}

The link consists of two single-mode dark fibres connecting laboratories at the University of Calgary (U of C) and the Southern Alberta Institute of Technology (SAIT), see \fref{fig5_1}. The fibres, which we refer to as channel 1 and channel 2, run through tunnels on the two campuses, and are buried or run through train tunnels in between the two institutions. They feature insertion loss of 7.8 dB and 6.5 dB, respectively. The fibre length is 12.4 km while the straight-line distance between the two laboratories is 3.3 km. A 1300 nm optical time-domain reflectometer (OTDR) with a 1 km dead zone eliminator was used to characterize the installed fibres. Figure \ref{fig5_2} shows the measured OTDR traces. The figure clearly shows that the last several kilometers of fibre have bad connections, which result in high transmission loss in our system. The peaks at the distance of 1 km are induced by the core diameter mismatch between the tested fibre and the dead zone eliminator, where the latter one is a multi-mode fibre.

\subsection{Polarization Transformation}
\begin{figure}
  \begin{center}
    \subfigure[]
      {
        \includegraphics{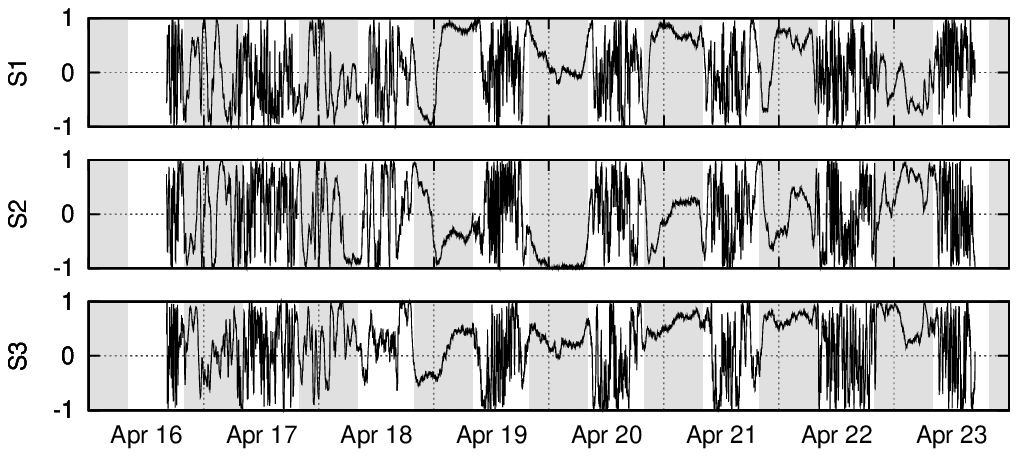}
        \label{fig5_3a}
      }
    \subfigure[]
      {
        \includegraphics{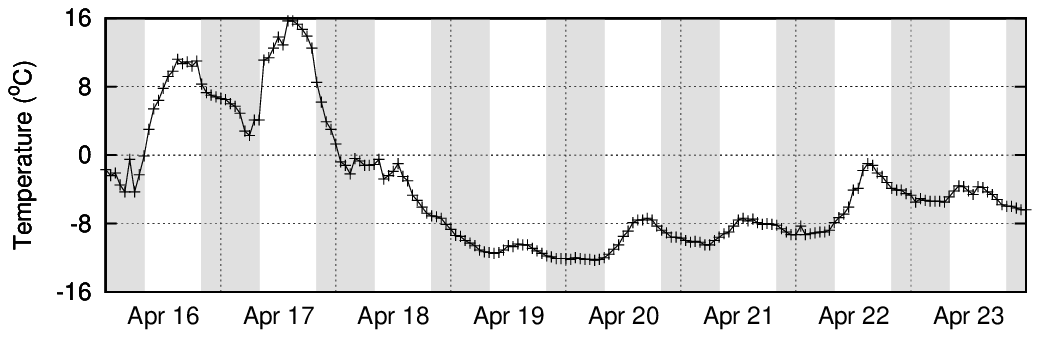}
        \label{fig5_3b}
      }
    \subfigure[]
      {
        \includegraphics{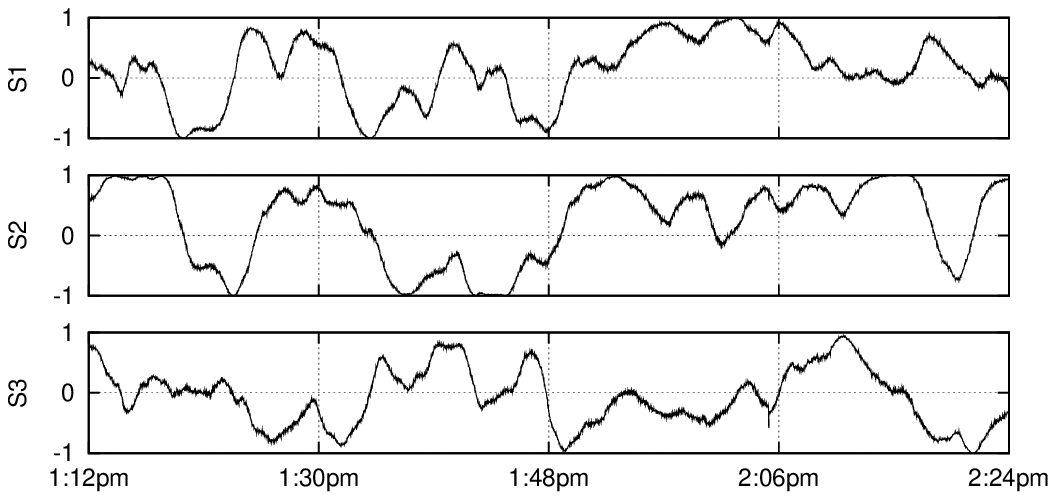}
        \label{fig5_3c}
      }
    \caption{(a)Time evolution of Stokes parameters during a full week in 2008. The shaded regions indicate night-time from 8:00 p.m. to 8:00 a.m. (b) Temperature curve for Calgary. (c) Zoom of (a) around April 19, lunch time.}
  \end{center}
\end{figure}

We experimentally studied the time evolution of polarization in the installed fibre. In the experiment, a stable polarized light source was launched into the fibre link, where channel 1 and channel 2 were looped at SAIT. We used a polarimeter to record Stokes parameters of the output light every second. Figure \ref{fig5_3a} presents the results of one-week of continuous monitoring from April 16, 2008 to April 24, 2008. Figure \ref{fig5_3b} shows the temperature curve for the Calgary Airport during the measurement (data from Canada Environment Weather Office). Comparing figure \ref{fig5_3a} and figure \ref{fig5_3b}, we observe a clear correlation between the variation of temperature and the fluctuation of polarization. This phenomenon is particularly obvious for the measurement from April 19 to April 23, where we observe small polarization variation during night, and much more pronounced variations during day-time. Figure \ref{fig5_3c} is a zoom-in of the measurement on April 19 (around lunch time), where particularly rapid polarization fluctuations are observed. Even for this case, we find that the polarization is stable on a time scale of tens of seconds. This sets an upper limit to the duration of quantum data between consecutive stabilization cycles.

\section{Field Tests \label{sec6}}

\subsection{Setup}

\begin{figure}
  \begin{center}
    \includegraphics[width=13cm]{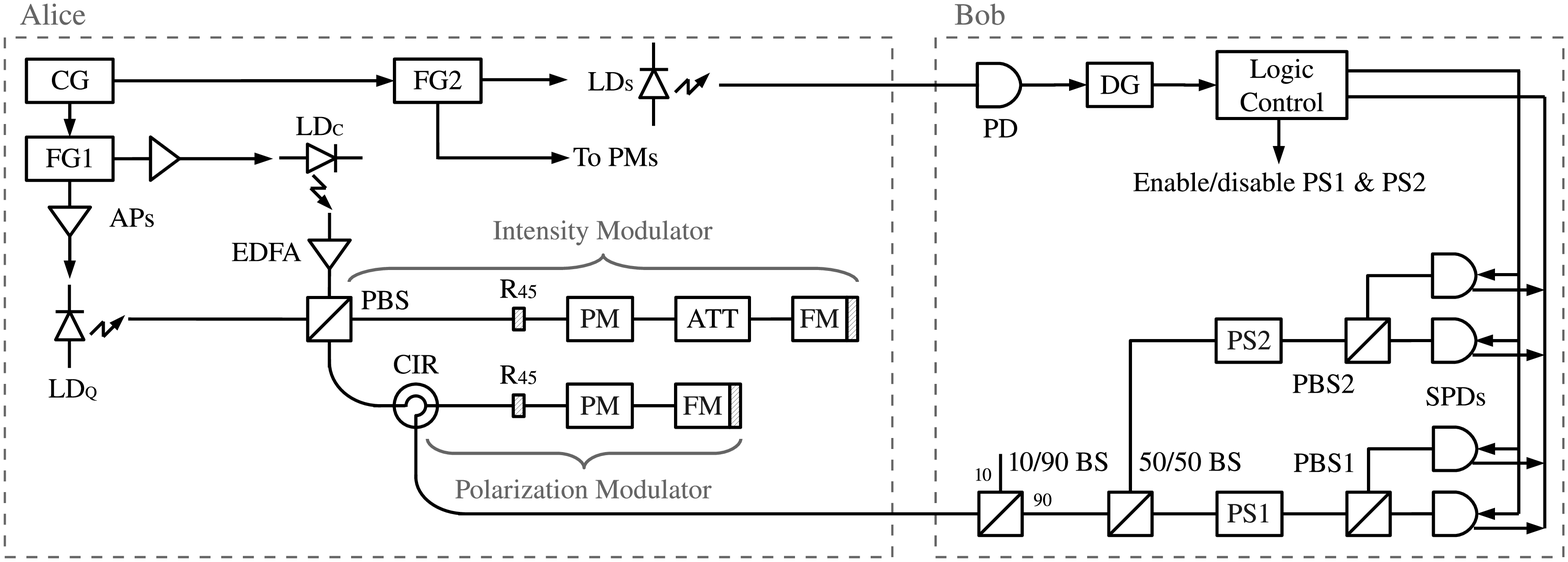}
    \caption[Caption short.]{Schematics of the QKD setup. \label{fig6_1}}
  \end{center}
\end{figure}

A schematic of the complete experimental setup is shown in \fref{fig6_1}. A 10 GS/s function generator (FG1) with two independent outputs drives the quantum laser diode (LD$_\mathrm{Q}$) and the classical laser diode (LD$_\mathrm{C}$) via broadband RF amplifiers (APs). Both laser diodes produce horizontally polarized optical pulses with a duration of 500~ps and a repetition rate of 50 MHz. By adjusting the temperature, we could closely match the spectral properties of the two laser diodes.  We obtained center wavelengths of 1548.07 nm and 1548.11 nm, and spectral widths (FWHM) of 0.214~nm and 0.224~nm for LD$_\mathrm{Q}$  and LD$_\mathrm{C}$, respectively. This is important to ensure that the polarization transformation sensed by means of the C-frames (generated with LD$_\mathrm{C}$) equals the one experienced by the quantum data (generated with LD$_\mathrm{Q}$).

The pulses from LD$_\mathrm{Q}$, eventually encoding quantum data at different mean photon numbers, propagate through a two-by-two polarizing beam splitter (PBS) and enter the intensity modulator, which is described in detail in \sref{sec4}. To reduce their energy to the single-photon level, a fixed optical attenuator (ATT) is placed between the Faraday mirror (FM) and the phase modulator (PM). Birefringence and polarization dependent loss of the attenuator are automatically compensated by the Faraday effect and therefore a stable attenuation is achieved. At the output of the PBS, the now vertically polarized weak laser pulses are combined with the horizontally polarized strong pulses from LD$_\mathrm{C}$, which encode the C-frame, to form a complete Q-frame. Quantum and classical data is then sent through the polarization modulator, which is also presented in \sref{sec4}. The intensity and the polarization modulator are driven by a function generator (FG2) with a pulse width of 4 ns. Note that the polarization maintaining circulator (CIR) that is part of the polarization modulator only allows horizontally polarized light to enter, while the pulses from LD$_\mathrm{C}$ and LD$_\mathrm{Q}$ impinge with orthogonal polarization. Therefore, we aligned the axes of the polarization maintaining fibre at the output of the PBS at 45 degrees with respect to the axes of the polarization maintaining fibre at the input of the circulator. This alignment makes the circulator work with both directions of polarization, yet, at the expense of 3 dB loss. Finally, the polarization modulated data is  forwarded to Bob through fibre channel 2.

Alice's electronic equipment is synchronized using a clock signal at 10 MHz from a clock generator (CG). Using a function generator, a laser diode (LD$_\mathrm{S}$), a photodiode (PD), and a delay generator (DG), the clock signal (reduced to 1 MHz) is also transmitted to Bob, where it provides trigger signals for the single photon detectors, synchronized with the arrival time of the quantum data.

At Bob's side, 90\% of the optical power encoded into each Q-frame is transmitted through a 10/90 beam splitter and is then equally divided by a 50/50 beam splitter. For each part, the C-frames are sensed by a polarization stabilizer (PS, from General Photonics) to compensate for the polarization change in the transmission line, and the quantum data is detected by a measurement module consisting of a PBS and two InGaAs based single photon detectors (SPDs). The SPDs are triggered at 1 MHz, and operated with a gate width of 5 ns, a deadtime of 10 ${\mu}s$ and a quantum efficiency of 10\%.

In principle, the length of a C-frame is determined by the response time of the polarization stabilizer, which is 18 ms. However, due to the small duty cycle of the classical pulse sequence in the current implementation and the low transmission of the fibre link, the average power of the C-frame is below the detection threshold of the polarization stabilizer. To resolve this problem, we placed a polarization maintaining erbium doped fibre amplifier (EDFA) between LD$_\mathrm{C}$ and the PBS. The EDFA is turned off after each C-frame to avoid flooding the SPDs at Bob's with photons from amplified spontaneous emission. While the turn-off time is only tens of milliseconds (consistent with the radiative lifetime of population in the upper laser level), we found the turn-on time of the EDFA to be as long as 3 seconds, resulting in 5-second long C-frames. The length of quantum data is set to 2 seconds, according to the ``worst-case" polarization stability of the fibre link, which is discussed in \sref{sec5}. From this, we find that our setup currently limits the time for quantum key distribution to 30\% of the operation time. Note, however, that the duty cycle of the classical pulse sequence can easily be increased by several orders of magnitude. In this case the duration of a C-frame would be limited by the response time of the polarization stabilizer, and the time for QKD could exceed 99\% of the system's operation time.

\subsection{Measurements}

We performed a variety of measurements to assess the performance of our QKD system. For \textit{2-detector measurements}, Alice repetitively creates  sequences of Q-frames with polarizations HH, HL, HV, HR, LH, LL, LV, LR, VH, VL, VV, VR, RH, RL, RV, and RR. The first letter indicates the polarization of the C-frame and the second one indicates that of the quantum data. Bob uses one measurement module to process the frames. The polarization stabilizer compensates the polarization transformation in the quantum channel for states belonging to the basis indicated by the first letter, i.e. linear or circular. For \textit{4-detector measurements}, Alice modulates the polarization of the Q-frames in the more complicated order of HH, RH, VH, LH, RH, HH, LH, VH, HR, RR, VR, LR, RR, HR, LR, VR, HV, RV, VV, LV, RV, HV, LV, VV, HL, RL, VL, LL, RL, HL, LL, VL. Bob uses two measurement modules to process the Q-frames. The polarization stabilizer of one module is always activated for odd frame numbers, and that of the other module is always activated for even frame numbers (see \fref{fig2_1}). In this way, the two measurement modules compensate polarization transformation for states encoded in the linear, or the circular basis, respectively. We collect the number of trigger events and counts for all single photon detectors for each combination of polarization states and different mean number of photons per qubit. This allows calculating average quantum bit error rates (QBER) and key generation probabilities (KGP), where the KGP is defined as the probability of generating a sifted key bit from a qubit encoded into a weak signal state when Alice and Bob use the same basis:

\begin{eqnarray}
  \eqalign{
    QBER &= \frac{P_\mathrm{wrong}}{P_\mathrm{wrong}+P_\mathrm{correct}}, \\
    KGP &= P_\mathrm{correct} + P_\mathrm{wrong}.
  }
\end{eqnarray}

\noindent The probabilities for correct ($P_\mathrm{correct}$) and wrong sifted key bits ($P_\mathrm{wrong}$) are obtained from experimental data by dividing the number of correct, or wrong, detection events by the number of trigger events. We assume that the probability for both detectors to click simultaneously can be ignored. In our setup, it was at least four orders of magnitude smaller compared to the probability for a single click. Note that in an actual implementation simultaneous clicks in two or more detectors have to be replaced by a randomly selected detection event \cite{Lutkenhaus1999_PRA, Lutkenhaus2000_PRA}.

Assuming that the photon number per laser pulse satisfies a Poissonian distribution, $P_\mathrm{correct}$ and $P_\mathrm{wrong}$ can be calculated using

\begin{eqnarray}
  \eqalign{
    P_\mathrm{correct} & = 1 - \sum_{n=0}^\infty \frac{\mu^n e^{-\mu}}{n!}(1-\frac{Y_0}{2})(1-t \eta a)^n \\
    & = 1-(1-\frac{Y_0}{2})e^{-\mu t \eta a}\\
    P_\mathrm{wrong} & = 1 - \sum_{n=0}^\infty \frac{\mu^n e^{-\mu}}{n!}(1-\frac{Y_{0}}{2})\big(1-t \eta (1-a)\big)^n \\
    & = 1-(1-\frac{Y_0}{2})e^{-\mu t \eta (1-a)}.
  }
  \label{eq6_2}
\end{eqnarray}

\noindent $Y_0/2$ is the probability for a detector click without Alice sending a photon, which includes detection events due to dark counts and stray photons. We found this probability in our setup to be equivalent to the dark count rate. $\mu$ is the average photon number of the weak pulses at Alice's output, $t$ is the overall transmission, which includes the fibre link and Bob's optical components, and $\eta$ is the quantum efficiency of the single photon detectors. Finally, $a$ describes the polarization extinction ratio of the PBS, i.e. the probability for a horizontally polarized photon to be transmitted through the PBS, normalized to the probability to exit.

\begin{figure}
  \begin{center}
    \includegraphics{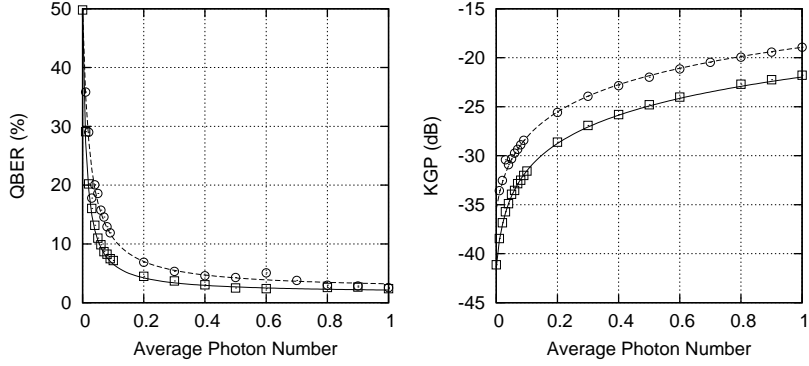}
    \caption[Caption short.]{Average QBER and KGP (in dB) as a function of the mean photon number per weak laser pulse used to encode the polarization qubits. The squares and circles indicate the experimental results for the 2-detector and 4-detector measurements, respectively, and the solid and dashed lines are the corresponding theoretical predictions (no fit). Error bars (corresponding to one standard deviation) are smaller than the size of each experimental data point. }
    \label{fig6_2}
  \end{center}
\end{figure}

The experimental results of the measurements are summarized in \fref{fig6_2}, together with the theoretical predictions. Note that all parameters required to calculate the QBER and the KGP have been obtained through independent measurements. We see that the experimental values match the theoretical calculations very well. We also find that the average QBER of the 4-detector measurement is larger than that of the 2-detector measurement at the same mean photon number. This is due to an increased dark count probability of the two additional SPDs, and slightly worse alignment of the polarization stabilizer in the second measurement module. Furthermore, the 4-detector measurement features a higher KGP as no qubits are lost at the 50/50 beam splitter. The individual data of the 4-detector measurement with an average photon number of 0.5 photons per pulse is listed in \tref{tab6_1}.

\begin{table}
  \caption{Results of the 4-detector measurement with an average photon number of 0.5 photons per pulse, where $pol$ indicates the polarizations of the Q-frames, $det$ and $trg$ are the number of photon detections and trigger events recorded by the single photon detectors, and $prob$ is the detection probability (in dB).\label{tab6_1}}
  \begin{indented}
  \item[]\begin{tabular}{c|rrc|rrc}
    \br
    ~ & \multicolumn{3}{c}{SPD$_\mathrm{1}$} & \multicolumn{3}{|c}{SPD$_\mathrm{2}$} \\
    \multicolumn{1}{c}{pol} &  \multicolumn{1}{|c}{det} & \multicolumn{1}{c}{trg} & \multicolumn{1}{c}{prob (dB)} &  \multicolumn{1}{|c}{det} & \multicolumn{1}{c}{trg} & \multicolumn{1}{c}{prob (dB)} \\
    \mr
    HH &  1,569 & 13,254,716 & $-$39.27 & 37,639 & 12,504,218 & $-$25.21 \\
    HV & 39,642 & 13,381,789 & $-$25.28 &  1,922 & 13,385,381 & $-$38.43 \\
    RR &  1,243 & 13,160,359 & $-$40.25 & 35,711 & 12,443,131 & $-$25.42 \\
    RL & 41,856 & 13,521,618 & $-$25.09 &  1,979 & 13,505,244 & $-$38.34 \\
    VH & 42,567 & 12,853,157 & $-$24.80 &    950 & 12,863,193 & $-$41.32 \\
    VV &  1,569 & 13,183,509 & $-$39.24 & 34,723 & 12,454,406 & $-$25.55 \\
    LL & 41,800 & 13,514,989 & $-$25.10 &  1,841 & 13,114,840 & $-$38.53 \\
    LR &    959 & 10,908,918 & $-$40.56 & 30,270 & 10,273,810 & $-$25.31 \\
    \br
    \br
    ~ & \multicolumn{3}{c}{SPD$_\mathrm{3}$} & \multicolumn{3}{|c}{SPD$_\mathrm{4}$} \\
    \multicolumn{1}{c}{pol} &  \multicolumn{1}{|c}{det} & \multicolumn{1}{c}{trg} & \multicolumn{1}{c}{prob (dB)} &  \multicolumn{1}{|c}{det} & \multicolumn{1}{c}{trg} & \multicolumn{1}{c}{prob (dB)} \\
    \mr
    HH &  37,577 & 12,468,543 & $-25.21$ &  1,050 & 12,416,198 & $-$40.73 \\
    HV &   2,121 & 12,145,604 & $-$37.58 & 35,843 & 11,410,147 & $-$25.03 \\
    RR &  35,954 & 12,409,015 & $-$25.38 &  1,605 & 12,351,662 & $-$38.86 \\
    RL &   3,222 & 12,253,689 & $-$35.80 & 36,378 & 11,541,004 & $-$25.01 \\
    VH &   2,410 & 12,817,201 & $-$37.26 & 39,290 & 12,046,285 & $-$24.86 \\
    VV &  36,215 & 12,403,829 & $-$25.35 &   925  & 12,355,805 & $-$41.26 \\
    LL &   2,751 & 12,270,811 & $-$36.49 & 36,547 & 11,534,262 & $-$24.99 \\
    LR &  29,988 & 10,247,919 & $-$25.34 &  1,149 & 10,193,024 & $-$39.48 \\
    \br
  \end{tabular}
  \end{indented}
\end{table}

\subsection{Long-Term Stability of the System}

\begin{figure}
  \begin{center}
    \subfigure[]
      {
        \includegraphics[width=5.5cm]{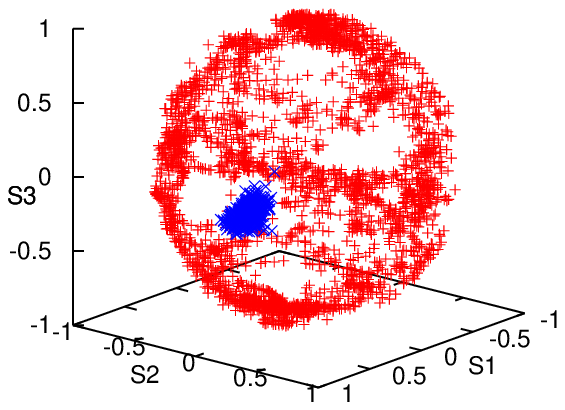}
        \label{fig6_3a}
      }
    \subfigure[]
      {
        \includegraphics[width=5.5cm]{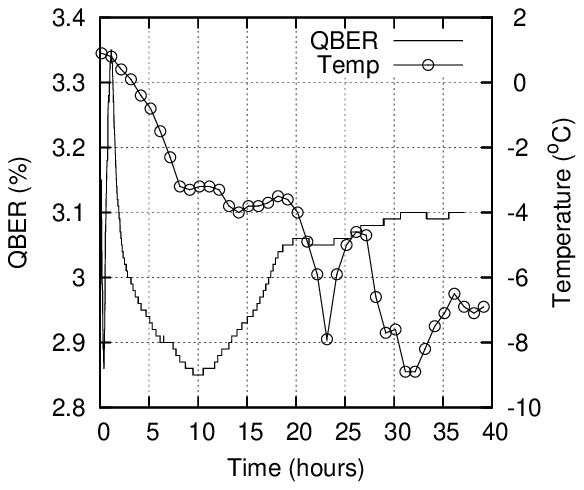}
        \label{fig6_3b}
      }
    \caption{Results of the long-term measurement. (a) Stokes vectors of C-frames with (blue points) and without (red points) polarization stabilization. (b) Average QBER and temperature for the same time interval as a function of time.}
  \end{center}
\end{figure}

To study the stability of the system, we performed a long time measurement over 37 hours. In the measurement, Alice sends qubits encoded into weak laser pulses with an average photon number of 0.5, and Bob implements a 2-detector measurement using measurement module one. At the end of each C-frame, i.e. after stabilization, Bob records the polarization of the C-frame with PS1. Meanwhile, the polarization stabilizer (PS2) in the second measurement module monitors the polarization of the C-frame without polarization control. In \fref{fig6_3a}, the red points indicate the Stokes vectors of the classical pulses measured by PS2, which are randomly distributed on the surface of the Poincar\'{e} sphere due to the time-varying polarization transformation in the transmission line. The blue points depict the measurements made by PS1, i.e. after polarization control. Even though the result slightly deviates from a single spot, which is expected in the ideal case, it clearly demonstrates the good long-term stability of our QKD system.

For a more quantitative analysis, we also recorded the evolution of the QBER over the same time interval, see \fref{fig6_3b}. The temperature curve for the Calgary Airport (data from Canada Environment Weather Office) is shown as well. The QBER varies between 2.85\% and 3.35\% in over 35 hours, and the variation is less than 0.1\% in the last 15 hours.

\section{Security Issues\label{sec7}}

For any cryptographic system, be it of quantum or classical nature, it is important to carefully analyze the actual implementation for weak points that may compromise its principle security. Applied to quantum key distribution, these include deficiencies in the preparation of quantum data at Alice's that can be exploited by an eavesdropper to gain information about the sifted key. We refer to these kind of attacks as \textit{quantum state attacks}. Furthermore, Eve may also attempt to actively sense the classical devices that create or measure the quantum data, or try to actively impact on the interaction between quantum and classical systems to influence the outcomes of measurements. We refer to these kind of attacks as \textit{classical system attacks}.

Note that, once the deficiencies are found, it may be possible to eliminate them by devising a better optical setup, or to remove the corresponding amount of information that Eve may have obtained through additional privacy amplification \cite{priv_amp}. Yet, we point out that loopholes may also arise from a careless implementation of privacy amplification, e.g. improper choice of Hash function, or of insufficient authentication of the classical channel. Finally, the size of the error corrected key has to be considered when calculating the appropriate amount of privacy amplification, i.e. to distil a secure key \cite{Scarani2008b, Hayashi2007}.

In the following, we will briefly discuss our current optical setup in view of such weak points. Yet, a complete security analysis of our system is beyond the scope of this article, which is the introduction of quantum frames. Note that the existence of loopholes in a particular QKD setup breaks the unconditional security of this particular system, but does not disprove that QKD can, in principle, be information theoretic secure.

\subsection{Quantum State Attacks}

The use of attenuated laser pulses, as opposed to pairs of entangled photons \cite{Gisin2002}, entails the possibility that non-orthogonal qubit states (here encoded into the polarization degrees of freedom) may become distinguishable when taking into account other degrees of freedom needed to fully describe the quantum data, e.g. frequency, temporal modes, or transverse modes. Obviously, in this case, the security offered by QKD would break down. We refer to these attacks as \textit{quantum side channel attacks}. Furthermore, as the number of photons in the attenuated laser pulses is described by a Poissonian distribution, it may be possible for an eavesdropper to gain information based on \textit{photon-number-splitting (PNS) attacks}.

\subsubsection*{Attacks Exploiting Quantum Side Channels:}

In our QKD system, all four qubit states are produced by the same laser diode, which is triggered independently of the subsequent action of the polarization or intensity modulators. Together with the polarization independent spectral transmission of both modulators and the attenuator, due to the use of the Faraday mirrors, this ensures that correlation between polarization state and spectrum or temporal mode do not exist. However, we recall that the circulator (CIR) at the output of the polarization modulator adds basis dependent polarization mode dispersion, which manifests as a basis dependent QBER. This may induce detectable temporal broadening of the photonic wavepackets, i.e. may partially reveal the basis used for encoding the qubit. The circulator will be replaced in a future, improved setup.

Furthermore, as the entire setup is built with (transverse) single mode optical fibres, correlation between polarization states and transverse modes, which may be present in a free space system, are ruled out.

\subsubsection*{PNS Attacks and Decoy States:\label{Security_PNS}}

The use of faint laser pulses makes our system principally susceptible to photon-number-splitting (PNS) attacks, which were first mentioned in \cite{Gisin1995} and have been analyzed thoroughly in \cite{Dusek1999,Brassard2000}. A possibility to remove the threat of the PNS attack is the use of so-called decoy states \cite{Hwang2003,Ma2005,Wang2005}. This allows establishing a conservative lower bound for the key that can be created from single photons emitted at Alice's, i.e. key that was not subject to the PNS attack. As described before, our setup has been devised to allow for the implementation of decoy states. In the following we will examine experimentally the accuracy with which the decoy state method allows bounding the size of the secret key. 

With the GLLP method the secure key rate per emitted faint pulse with mean photon number of $\mu$ is given by \cite{GLLP}

\begin{equation}
  S \geq
  \frac{1}{2} \big[
  Q_\mathrm{1} (1 - H_\mathrm{2}(E_\mathrm{1})) -
  Q_\mathrm{\mu} f(E_\mathrm{\mu}) H_2(E_\mathrm{\mu})
  \big]
  \label{eq7_1}
\end{equation}

\noindent where the factor $1/2$ accounts for basis reconciliation, $H_2(x)=-xlog_2(x)-(1-x)log_2(1-x)$ denotes the Shannon entropy, $Q_1$, $Q_{\mu}$, $E_1$ and $E_{\mu}$ specify the gains and error rates of signal states and single photons, respectively, and $f(E_{\mu})$ is the error correction efficiency which is assumed to be 1.22 \cite{Brassard94}.

In the first analysis, we assume that no PNS attack took place during the measurement, which is a reasonable assumption. Using equations \eref{eq6_2}, we can estimate the gain and error rate for signal states with mean photon number $\mu$: 

\begin{equation}
  \eqalign{
    Q_{\mu} & = P_\mathrm{correct}(\mu)+P_\mathrm{wrong}(\mu) \\
    & = 2-(1-Y_{0}/2)(e^{-\mu t \eta a}+e^{- \mu t \eta (1-a)}) \\
    E_{\mu} & = \frac{P_\mathrm{wrong}(\mu)}{P_\mathrm{correct}(\mu)+P_\mathrm{wrong}(\mu)} \\
    & = \frac{1-(1-Y_{0}/2)e^{-\mu t \eta a}}{2-(1-Y_{0}/2)(e^{-\mu t \eta a}+e^{-\mu t \eta (1-a)})}.
  } \label{eq7_3}
\end{equation}

\noindent Similarly, the gain and error rate for single photon pulses are given by

\begin{eqnarray}
  \eqalign{
    Q_\mathrm{1} & = \mu e^{-\mu} \big( 2 - \big( 1 - Y_0/2 \big) (2-t \eta) \big) \\
    E_\mathrm{1} & = \frac{1 - \big(1 - Y_0/2)(1 - (1-a)t \eta)} {2 - \big( 1 - Y_0/2 \big)(2-t \eta) }.
  } \label{eq7_5}
\end{eqnarray}

\begin{figure}
  \begin{center}
    \includegraphics{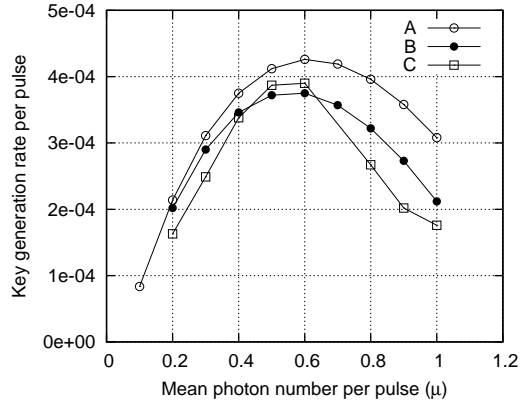}
    \caption{Comparison of secret key rates versus mean number of photons in the signal states. Curve A is the secret key rate calculated from the fraction of single photons emitted at Alice's and assuming fair loss (i.e. assuming it is known that all loss is of technological origin and that there is no PNS attack). Curve B shows the secret key rate calculated via the decoy state method (using decoy states with mean photon number of 0.1 and vacuum states) and assuming fair loss. Curve C is the secret key rate obtained via the decoy state method using experimental data. All calculations assume an infinite sifted key length. \label{fig7_1}}
    \end{center}
\end{figure}

\noindent Using equation \eref{eq7_1}, \eref{eq7_3} and \eref{eq7_5} and taking into account the measured values for $t$, $\eta$, $a$ and $Y_0/2$,  we can calculate the secret key rate for different $\mu$, see curve A of \fref{fig7_1}.

In the second analysis, which again relies on the assumption of fair loss, we use equation \eref{eq7_3} to calculate the gains and error rates for the signal state with mean photon number $\mu$ and the decoy state with mean photon number $\nu$ of 0.1. To calculate the gain and error rate for single photon pulses, we use equations (34), (35) and (37) from \cite{Ma2005}:

\begin{equation}
  \eqalign{
    Q_\mathrm{1} \geq Q_\mathrm{1}^\mathrm{\nu, 0} = \frac
    {\mu^\mathrm{2} e^\mathrm{-\mu}}{\mu \nu - \nu^\mathrm{2}} \Big(
    Q_\mathrm{\nu} e^\mathrm{\nu} - Q_\mathrm{\mu} e^\mathrm{\mu}
    \frac{\nu^\mathrm{2}}{\mu^\mathrm{2}}-
    \frac{\mu^\mathrm{2}-\nu^\mathrm{2}}{\mu^\mathrm{2}} Y_\mathrm{0}
    \Big) \\
    e_\mathrm{1} \leq e_\mathrm{1}^\mathrm{\nu,0}= \frac{E_\mathrm{\nu}
      Q_\mathrm{\nu} e^\mathrm{\nu} - e_\mathrm{0}
      Y_\mathrm{0}}{Y_\mathrm{1}^\mathrm{L, \nu,0} \nu} \\
    Y_\mathrm{1} \geq Y_\mathrm{1}^\mathrm{L,\nu,0} =
    \frac{\mu}{\mu\nu-\nu^\mathrm{2}} \Big( Q_\mathrm{\nu}
    e^\mathrm{\nu}-
    Q_\mathrm{\mu}e^\mathrm{\mu}\frac{\nu^\mathrm{2}}{\mu^\mathrm{2}}-\frac{\mu^\mathrm{2}-\nu^\mathrm{2}}{\mu^\mathrm{2}}Y_\mathrm{0}
    \Big).
  }
  \label{eq7_6}
\end{equation}

\noindent The resulting secret key rate follows from equation \eref{eq7_1}. It is shown in curve B of \fref{fig7_1}.

Finally, we calculate the secret key rate using the experimentally measured gain and error rates for signal and decoy states, as opposed to the previous case where they were calculated. The gain and error rate for single photons are estimated as before using equations (\ref{eq7_6}). The result is plotted in curve C of \fref{fig7_1}. Note that the measurement does not rely on the fair loss assumption.

Comparing the three different curves, we find that the rates estimated from the decoy state method (curves B and C) is somewhat smaller than the one plotted in curve A. This is natural as the decoy state method with decoy states of finite photon mean number only yields a conservative lower bound \cite{Ma2005}. As an example, for $\mu=0.6$, we find the secret key rate (curve B and C) to be roughly 10\% worse than the secret key rate given in curve A. We also find a reasonably good agreement between the rates estimated and measured using the decoy state method (curves B and C, respectively). We attribute the remaining discrepancy to a systematic error in the estimation of the single photon gain $Q_\mathrm{1}$, resulting from a slightly wrong estimation of the transmission in the link, quantum efficiency of the detectors, or error rate due to wrongly received photons. Factors like fluctuations in the mean photon number could also have an effect. This systematic error also affects the estimation of the single photon error rate $E_\mathrm{1}$. Furthermore, curve B and C show that the secret key rate in our QKD system is maximized for signal states with a mean number of photons of $\mu \approx 0.6$. This value agrees with estimations in \cite{Ma2005} when taking into account the actual values for dark count rates, transmission, detector quantum efficiency, and error rate caused by wrongly received photons. Indeed, we calculate $\mu_{opt} = 0.62$, in very good agreement with our experimental results.

To finish this discussion, we emphasize that the secret key rate in an actual implementation of an information-theoretic secure QKD session must be calculated using the decoy state method used in the third analysis and must not rely on assumptions about fair loss in the quantum channel.

\subsubsection*{Other deficiencies:}

We have noted that each faint pulse that encodes a qubit is preceded by another faint pulse, originating from a reflection on the PBS that is part of the intensity modulator (see section \ref{sec4}). Note that the number of photons in both pulses is comparable. Obviously, for our assessment of the eavesdropper's information to be correct, we have to make sure that this pulse, which also transits through the polarization modulator, does not encode any polarization information. Therefore, we have carefully adjusted the electrical trigger signal for the polarization modulator such that it only acts on the ``real" faint pulse, and not on the spurious one.

\subsection{Classical System Attacks}

\subsubsection*{Trojan Horse Attacks:}

As in any QKD system, regardless whether it employs one-way or two-way quantum communication, appropriate measures have to be implemented to protect against Trojan-Horse attacks \cite{Gisin2006}. In these attacks, the eavesdropper injects light through the optical fibre into Alice's or Bob's preparation or measurement device, respectively, and analyzes the back reflection, which may reveal information about the quantum state created at Alice's or the measurement basis to be used at Bob's. In both cases, the security of the key distribution would be compromised as Eve either knows the state, or knows in which basis to perform an intercept resend attack without creating errors. In our QKD system, given the static setup at Bob's, Trojan Horse attacks have to be considered only at Alice's. Towards this end, a polarization independent optical isolator and a spectral filter that absorbs all wavelengths not blocked by the isolator should be placed at the output of Alice's.

\subsubsection*{Time-shift attacks:}

In a time-shift attack \cite{Makarov2006,Lamas2007,Zhao2008} the eavesdropper exploits the fact that the detection efficiency of different detectors may, for a given arrival time of a photon, be different. It may thus be possible for an eavesdropper to bias the detection probabilities by actively time-shifting the arrival time of photons and thereby acquire information for each photon if it was detected in a detector that codes for a bit value 0, or 1. This attack, which is possible in our current system, can be overcome if Bob randomly rotates the polarization state of each incoming qubit by 0 or $\pi/2$, thereby de-correlating a detection in a particular detector with a particular bit value. This can be done by placing a rapidly variable $\lambda/2$ waveplate in between the polarization stabilizer and the PBSs, at the expense of rendering Bob's setup ``active", i.e. vulnerable to Trojan Horse attacks (which then have to be protected against, as discussed above).

\section{Classical Post-Processing\label{sec8}}

\begin{figure}
\begin{center}
\includegraphics[width=20pc]{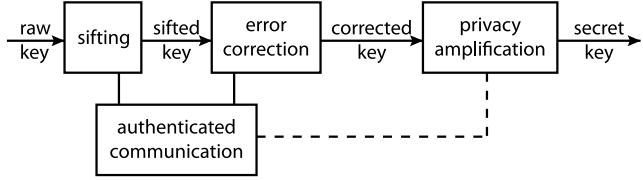}
\caption{Classical post-processing steps.\label{fig8_1}}
\end{center}
\end{figure}

Once the quantum part of the QKD protocol is finished, Alice and Bob must
perform a series of classical steps to go from the raw key to the secret
key used for encryption \cite{Gisin2002}. The steps required are shown in
\fref{fig8_1}.  In addition to sifting, error correction is used to ensure
that Alice and Bob have an identical key despite any errors that occur.
Privacy amplification is then used to eliminate any information Eve has
obtained about the key, whether through eavesdropping on the quantum
channel or on the classical communication used for error correction. These
steps must also make use of authenticated communication to prevent Eve
from performing a man-in-the-middle attack.  Of these steps, error
correction is expected to become the bottleneck in the QKD system once
higher raw key rates are achieved.  The Cascade protocol~\cite{cascade}
that was originally developed for QKD is not suitable for high key rates
as it requires many rounds of communication between Alice and Bob and is
computationally expensive~\cite{Pearson}.

\subsection{Low-Density Parity-Check Codes\label{sec:post_ldpc}}

Low-Density Parity-Check (LDPC) codes were originally developed by
Gallager in the 1960s~\cite{Gallager} for classical communications, but
their potential performance has only been recently been
discovered~\cite{MacKay}.  LDPC codes for QKD differ slightly from those
used in the classical case as the parity information is transmitted over a
separate classical channel~\cite{Pearson}.

A LDPC code is defined using an $m\times n$ parity check matrix, $H$,
consisting of zeros and ones.  While either Alice's or Bob's sifted key
may be considered the ``correct" key for the purpose of error correction,
this discussion will use Alice's sifted key, the $n$ bit column vector
$\balpha$, i.e. one-way, forward error correction.  Alice computes a
parity vector as follows:

\begin{equation}
\bi{p} = H\balpha\qquad(\mbox{mod } 2),
\label{eqn:parity_check}
\end{equation}

\noindent where the number of bits $m$ in the parity vector is lower
bounded by Shannon's noisy coding theorem; $m=n H_2(QBER)$ with Shannon
Entropy $H_2$. Thus, $p_i$ indicates whether the sifted key bits indicated
by the ones in the $i^\mathrm{th}$ row of $H$ contain an even ($p_i$=0) or
odd ($p_i$=1) number of ones.  Alice transmits $\bi{p}$ to Bob, whose task
it is to determine $\balpha$ using $H$, $\bi{p}$, his sifted key,
$\bbeta$, and an initial estimate of the QBER.  This estimate can be based
on a characterization of the quantum channel or on the QBER from previous
executions of the protocol.

In order to recover $\balpha$, Bob uses a process known as belief
propagation to refine his initial probabilities for the entries of
$\balpha$ based on $\bbeta$ and the QBER. Note that in the following
discussion, Bob has full knowledge of his key vector, $\bbeta$, but his
knowledge of the Alice's key vector, $\balpha$ is probabilistic.  For
example, suppose row $i$ of $H$ is a parity check on three bits received
by Bob, $\beta_1=1$, $\beta_2=1$, and $\beta_3=0$, where the expected QBER
is 10\% (chosen to prevent very small numbers in this example).  The
probability that a key bit $\alpha_j$ is zero or one based on the received
values and the QBER are denoted $P_0(j)$ and $P_1(j)$, respectively.  For
each of his bits $\beta_j$, Bob assumes that $\alpha_j=1$ and computes
$r_{\alpha_j=1}(i,j)$, which denotes the probability that the parity check
$i$ is satisfied ($p_i = \alpha_1+\alpha_2+\alpha_3\mbox{ }(\mbox{mod }
2)$) given this assumption.  Alternatively, $r_{\alpha_j=1}(i,j)$ may be
viewed as the probability that $\alpha_j=1$ given the value of $p_i$ and
what is known about the other bits of $\balpha$ involved in the
$i^\mathrm{th}$ parity check.  For example, $r_{\alpha_j=1}(i,1)$ may be
computed as follows:

\begin{equation}
\label{cases}
r_{\alpha_j=1}(i,1)=\cases{P_0(2) P_1(3)+P_1(2) P_0(3)&for $p_i=0$\\
P_0(2) P_0(3)+P_1(2) P_1(3)&for $p_i=1$.\\}
\label{eqn:r}
\end{equation}

As can be seen in \tref{tab:cnode}, the probability that the bits retain
their received value is high when $p_i=0$ since this is consistent with
the received values of $\bbeta$.  If instead $p_i=1$, a high probability
for bit flips is obtained since each row assumes that the received values
for the other bits are likely to be correct. This information is useful
when combined with the results of other parity checks.

\begin{table}
\caption{Results for $r_{\alpha_j=1}(i,j)$.\label{tab:cnode}}
\begin{indented}
\item[]\begin{tabular}{@{}llllll}
\br
$j$ & $\beta_j$ & $P_0(j)$ & $P_1(j)$ & $r_{\alpha_j=1}(i,j)$ for $p_i=0$
& $r_{\alpha_j=1}(i,j)$ for $p_i=1$\\
\mr
1 & 1 & 0.1 & 0.9 & $0.82$ & $0.18$\\
2 & 1 & 0.1 & 0.9 & $0.82$ & $0.18$\\
3 & 0 & 0.9 & 0.1 & $0.18$ & $0.82$\\
\br
\end{tabular}
\end{indented}
\end{table}

After doing these computations for each row of $H$, Bob uses the
information from all the parity checks involving a particular key bit
$\beta_j$ to compute new values of $P'_0(j)$ and $P'_1(j)$.  If the
$j^\mathrm{th}$ key bit is involved in three parity checks, Bob computes
$q_{\alpha_j=0}(j)$ and $q_{\alpha_j=1}(j)$, which represent the
probability that $\alpha_j$ is zero or one, respectively, based on
$\beta_j$ and the QBER, and that all parity checks involving $\alpha_j$
are satisfied:
\begin{eqnarray}
q_{\alpha_j=0}(j) & = & P_0(j) r_{\alpha_j=0}(1,j) r_{\alpha_j=0}(2,j)
r_{\alpha_j=0}(3,j)\label{eqn:q_0}\\
q_{\alpha_j=1}(j) & = & P_1(j) r_{\alpha_j=1}(1,j) r_{\alpha_j=1}(2,j)
r_{\alpha_j=1}(3,j)\label{eqn:q_1}
\end{eqnarray}

\noindent where $r_{\alpha_j=0}(i,j)=1-r_{\alpha_j=1}(i,j)$.  Since valid
results must be consistent with all parity checks, $P'_0(j)$ and $P'_1(j)$
are obtained by normalizing $q_{\alpha_j=0}(j)$ and $q_{\alpha_j=1}(j)$.
For example, consider $\beta_j=1$, implying $P_0(j)=0.1$ and $P_1(j)=0.9$
as shown in \tref{tab:mnode}.  Even if one parity check suggests there is
an error in this example, the confidence that $\beta_j=1$ (i.e. $\beta_j$
was received correctly) still increases.  With all three parity checks
suggesting a bit flip is necessary, a high confidence is obtained that the
received value of $\beta_j$ is incorrect.  With two parity checks
suggesting a bit flip is required, the result does not significantly
favour either result.

Bob can then select the most likely value for each bit to form $\bbeta'$,
and compute $\bi{p}'=H\bbeta'\mbox{ }(\mbox{mod } 2)$.  If
$\bi{p}'=\bi{p}$, the protocol is finished.  Otherwise, additional
iterations of the protocol are performed.  With the additional
modification that Bob also computes conditional probabilities, $P'_0(i,j)$
and $P'_1(i,j)$, to use in \eref{eqn:r} during subsequent iterations, this
procedure is generalized as the sum-product
algorithm~\cite{Pearson,MacKay}.

\fulltable{\label{tab:mnode}Results for $P'_0(j)$ and $P'_1(j)$ values.}
\br
$r_{\alpha_j=1}(1,j)$ & $r_{\alpha_j=1}(2,j)$ & $r_{\alpha_j=1}(3,j)$ &
$q_{\alpha_j=0}(j)$ & $q_{\alpha_j=1}(j)$ & $P'_0(j)$ & $P'_1(j)$\\
\mr
0.82 & 0.82 & 0.82 & 0.0006 & 0.4963 & 0.0012 & 0.9988\\
0.18 & 0.82 & 0.82 & 0.0027 & 0.1089 & 0.0238 & 0.9762\\
0.18 & 0.18 & 0.82 & 0.0121 & 0.0239 & 0.3361 & 0.6639\\
0.18 & 0.18 & 0.18 & 0.0551 & 0.0052 & 0.9131 & 0.0869\\
\br
\endfulltable

\subsection{Hardware LDPC Decoding\label{sec:post_fixed}}

Interest in LDPC codes stems not only from their potential to perform near
the Shannon limit.  Since the computations for each parity check and each
key bit are independent, the structure of the sum-product algorithm lends
itself to parallel computation.  This makes sum-product decoding of LDPC
codes well suited for high speed implementation in custom hardware or in
reconfigurable devices such as Field Programmable Gate Arrays
(FPGA)~\cite{Levine}.  However, floating-point computations are expensive
in terms of the amount of logic required.  Thus, it is desirable to
implement LDPC decoding using fixed-point arithmetic (equivalent to
integer arithmetic) with as few bits as possible to represent the values.
In initial simulations of fixed-point decoding, we found that the primary
obstacle for a small bit length was the very small values obtained for the
probabilities.  This problem manifested as ``divide by zero" errors during
the normalization since both $q_0(j)$ and $q_1(j)$ had rounded to zero.
We overcome this limitation by modifying the algorithm to set any
occurrences of zero in the $q(j)$ values to the smallest possible non-zero
value.

\begin{figure}
\begin{center}
\includegraphics[width=20pc]{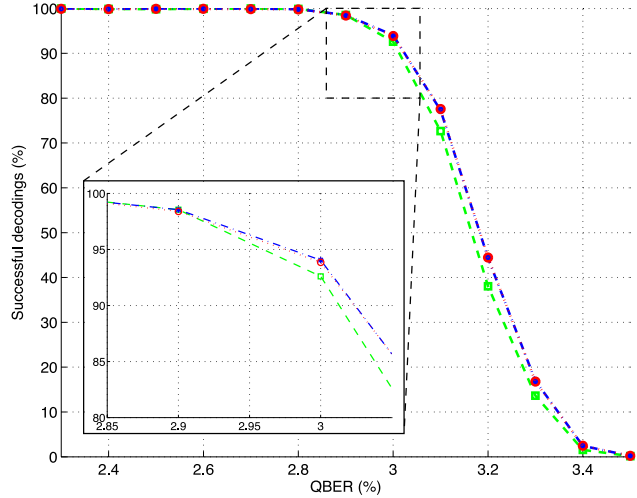}
\caption{Simulation results of the $1200\times4000$ LDPC code using 16-bit
fixed-point (\broken,\opensquare), 24-bit fixed-point (\chain,*), and
floating-point (\dotted,\opencircle).  The inset shows the region where
the performance begins to drop in more detail.\label{fig8_2}}
\end{center}
\end{figure}

A LDPC code was designed with a $1200\times4000$ parity check matrix using
parameters similar to~\cite{Pearson} (QBER=3\%, parity checks on 20 key
bits. Note that this QBER also reflects our experimental results, see
section \ref{sec5}).  It has been shown that having the key bits take part
in a variable number of parity checks results in better
performance~\cite{Luby}.  Thus, $H$ has a fixed number of ones in each
row, known as the row weight, and a variable number of ones in each
column, known as the column weight.  The method presented in~\cite{Luby}
was used to determine the column weights by applying a well known
optimization technique with the constraints ensuring that the design
criteria (QBER and code rate) are met.  In place of the arbitrary cost
function in~\cite{Luby}, we use a function reflecting the computational
complexity.  Our code was simulated over 40 iterations, with the number
being selected based on tests which showed very little improvement beyond
this point.  The results in \fref{fig8_2} show that 24-bit fixed-point and
floating-point have very similar decoding performance.

Using VHDL (a hardware description language) code generated in Matlab, we
are able to create code for parallel implementations of sum-product
decoding for arbitrary values of $H$.  While a RTL (Register Transfer
Level) simulation of the $1200\times4000$ LDPC code is possible, a fully
parallel implementation is not possible at this time.  A $60\times200$
LDPC code with a row weight of 12 that is capable of operating at 50MHz
was synthesized using the Artisan 3.0 logic cell library for 0.18
$\mathrm{\mu m}$ CMOS technology (several generations behind state of the
art).  This code uses 12-bit arithmetic and requires 46 clock cycles (0.92
$\mathrm{\mu s}$) per iteration of the algorithm.  Simulation results for
the performance of this code with a maximum of 40 iterations are given in
\tref{tab:200}.  The design contains 1860429 cells with a total cell area
of approximately 47.24 $\mathrm{mm}^2$.  Attempts to synthesize a larger
LDPC code using the current VHDL code have failed as the synthesis tool
does not have sufficient memory to complete the process.  The size of the
design also suggests that a $1200\times4000$ code would be impractical to
implement (as a comparison, a processor is typically on the order of 100
$\mathrm{mm}^2$, including interconnect).  However, larger codes are
preferred because they experience less variance from the mean QBER and
perform better relative to the Shannon limit.

\begin{table}
\caption{Simulation results for $60\times200$ LDPC decoding.\label{tab:200}}
\begin{indented}
\lineup
\item[]\begin{tabular}{@{}llll}
\br
QBER (\%) & Success rate (\%) & Mean iterations & Sifted key rate (Mb/s)\\
\mr
2.5 & 99.00 & \04.1070 & 52.9319\\
3.0 & 91.65 & \08.6785 & 25.0494\\
3.5 & 69.80 & 17.9455 & 12.1146\\
\br
\end{tabular}
\end{indented}
\end{table}

It is important to note that we obtained these results without using any
advanced techniques to reduce the size of the design.  More efficient
multiplier designs or the use of alternative number systems such as the
multidimensional logarithmic number system (MDLNS)~\cite{MDLNS} have the
potential reduce the hardware required to perform the computations.
Larger block sizes could also be achieved using the partially parallel
implementations proposed in~\cite{Sharon}, where efficient schedules are
used rather than updating all probabilities at once, reducing the number
of computations done in parallel while mitigating the cost in terms of the
run time.

\section{Conclusion and Outlook\label{sec9}}
We have proposed a novel, fibre-based QKD system employing polarization encoding and quantum frames, and have demonstrated in a long-term (37 hours) QKD proof-of-principle study that polarization information encoded in the classical control frames can indeed be used to stabilize unwanted qubit transformation in the quantum channel. All optical elements in our setup can be operated at Gbps rates, which is a first requirement for a future system delivering secret keys at Mbps. In order to remove another bottleneck towards a high rate system, we are investigating forward error correction based on Low-Density Parity-Check Codes~\cite{Gallager,MacKay}. Work on the implementation of a system that distributes a quantum key, building on the here presented proof-of-concept demonstration, is under way.

\small
\section*{Acknowledgments}
The authors gratefully acknowledge discussions with Xiongfeng Ma. This work is supported by General Dynamics Canada,
Alberta's Informatics Circle of Research Excellence (iCORE), the National Science and Engineering Research Council of Canada (NSERC), QuantumWorks, Canada Foundation for Innovation (CFI), Alberta Advanced Education and Technology (AET), CMC
Microsystems, and the Mexican Consejo Nacional de Ciencia y Tecnolog\'{\i}a (CONACYT).


\end{document}